\documentclass[journal,twocolumn]{IEEEtran}
\usepackage{array}
\usepackage{amsmath}
\usepackage{amssymb}
\usepackage{textpos}
\usepackage{subfigure}
\usepackage{enumerate}
\usepackage{lscape}
\usepackage{float}
\usepackage{pdfsync}
\usepackage{color}
\usepackage{url}
\usepackage{threeparttable}
\usepackage{multirow}
\usepackage{array,multirow}
\usepackage{booktabs}
\usepackage{tabularx}
\usepackage{amsthm}
\usepackage[font=small,labelfont=bf]{caption}
\usepackage{setspace}
\usepackage{graphicx}
\usepackage{epstopdf}
\usepackage{textcomp}
\usepackage{cite}
\usepackage{algorithmic}
\usepackage{array}
\usepackage{lettrine}
\usepackage{romannum}
\newtheorem{theorem}{Theorem}

\newtheorem{lem1}{Lemma}

\makeatletter
   \def\vhrulefill#1{\leavevmode\leaders\hrule\@height#1\hfill \kern\z@}
\makeatother
 \begin{document}
\newcommand{\E}{\mathbb{E}}
\title{Design and Performance Analysis of Secure  Multicasting Cooperative Protocol for Wireless Sensor Network Applications}
\author{Michael Atallah, and Georges Kaddoum,~\IEEEmembership{Member,~IEEE.}

\thanks{Michael Atallah, and Georges Kaddoum are with University of Qu´ebec, ETS, 1100 Notre-Dame west, H3C 1K3, Montreal, Canada (e-mail: michael.atallah.1@ens.etsmtl.ca;  georges.kaddoum@etsmtl.ca).}
\thanks{Manuscript received XXX, XX, 2018; revised XXX, XX, 2018.}}

{}
\maketitle
\begin{abstract}
This paper proposes a new security cooperative protocol, for dual phase amplify-and-forward large wireless sensor networks. In such a network, a portion of the $K$ relays can be potential  eavesdroppers. The source agrees to share with the destination a given channel state information (CSI) of a source-trusted relay-destination link to encode the message. Then, in the first hop, the source will use this CSI to map the right message to a certain sector while transmitting fake messages to the other sectors using sectoral transmission thanks to analog beamforming. In the second hop, the relays retransmit their received signals to the destination, using the distributed beamforming technique. We derived the secrecy outage probability and demonstrated that the probability of receiving the right encoded information by an untrustworthy relay is inversely proportional to the number of sectors. We also showed that the aggressive behavior of the cooperating untrusted relays is not effective compared to the case where each untrusted relay is trying to intercept the transmitted message individually.
\end{abstract}
\begin{IEEEkeywords}
Physical layer security, secrecy outage probability, amplify and forward, secrecy capacity.
\end{IEEEkeywords}
\section{Introduction}
\IEEEPARstart{I}{n} 
  wireless networks, nodes can join and leave frequently, which increases the risk of the malicious nodes that are penetrating the wireless network.
Therefore, the demand for security solutions in the physical layer is becoming more and more  essential. 
One of the important metrics that evaluate the security performance in the physical layer is the secrecy rate, which is the difference between the channel capacity of the legitimate links and the channel capacity of the illegitimate ones \cite{4626059}. 
Many techniques have been proposed to achieve a positive secrecy rate, such as cooperative jamming, multi-antenna scenarios, beamforming, game theory, and power allocation schemes \cite{7324413}. 
A wireless network could benefit from the new joining nodes, by using them as relays, or by treating them as potential eavesdroppers.
However, as shown in  
\cite{5508640}, taking advantage of these nodes and using them as relays could be more useful to the wireless network, from a security perspective, than treating them as eavesdroppers. 
The authors in \cite{7061484} studied the secrecy performance for the case of having multiple passive untrusted relays, where each passive untrusted relay is trying to intercept its received message individually. 
In  \cite{7463514}, the authors studied the secrecy capacity scaling with aggressive untrusted relays. {\color{black}{We define the aggressive behavior as when the untrusted relays are  cooperating between each other by sending their received messages to an external wiretapper. }}
Both \cite{7061484} and \cite{7463514} considered two transmission schemes, namely opportunistic relaying (OR) and distributed beamforming (DBF).
 They also demonstrated that  DBF outperforms  OR technique from a secrecy perspective.
In \cite{8113794}, a new location-based multicasting technique was proposed considering both passive and  aggressive untrusted relays behaviors.  It was shown that this technique enhances the security compared to \cite{7061484} and \cite{7463514}.  

On the other hand, the randomness of the channel has been exploited for different purposes, whether to enhance the reliability or to secure the communication system as it was used to generate
keys in \cite{1506252}.
Therefore, in this paper, we combine the channel randomness with multicasting transmission to propose a new location-based multicasting protocol in two-hops wireless sensor networks (WSN). The goal of  this protocol is to increase the security  of  these networks while taking into account that wireless sensor nodes have  limited capabilities.  
In the proposed protocol, the source and the destination share the channel state information (CSI) to map the source's transmission by sending the useful encoded message towards a specific sector, while sending other fake messages, similar to the useful one,  towards  the other sectors to confuse the eavesdroppers. 
We provide analytical expressions for  secrecy outage probability (SOP) for both passive and aggressive untrusted relays. 
Our numerical results show how our technique enhances the security performance and how immune it is against the aggressive behavior of the  untrusted relays. Finally, adopting such a security protocol  by
allowing a part of the nodes to forward fake messages is promising because of the availability of high number of cheap electronic sensors with limited computational capabilities.
\section{System Model and Problem Formulation}\label{systinf}
\begin{figure}
\centering
\includegraphics[trim =0 00 0 8, clip,width=3.2in]{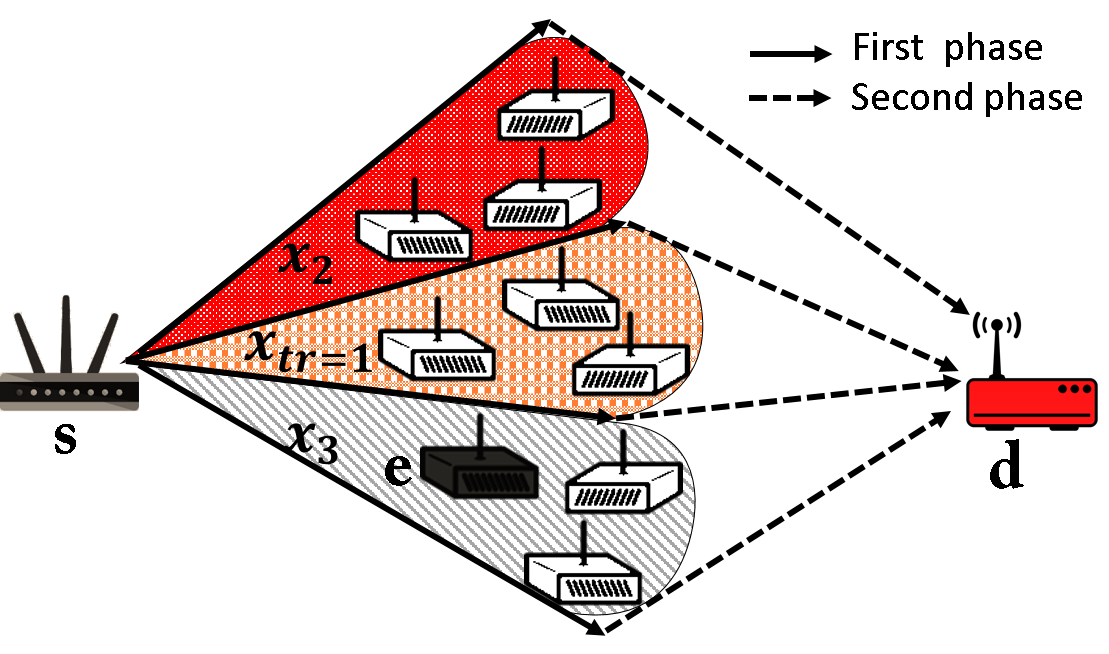}
\caption{In the 1st hop of each transmission, $s$ multicasts the useful message $x_{tr}$ and the fake ones $x_{i\ne {tr}}$'s towards $N$ sectors. In the 2nd hop, the $K$ relays retransmit their received messages towards $d$.  
}
\label{SystemModel}
\vspace{-12pt}
\end{figure}
Consider a multi-antennas source $s$, $K$ amplify-and-forward (AF) cooperative  relay sensor nodes with limited capabilities, and a destination $d$. Out of the $K$ relays, there are $U$ untrustworthy relays that could be  potential eavesdroppers.  Each relay is equipped with a single antenna and works in a half-duplex mode,  as shown in Fig.\ref{SystemModel}. It is assumed that there is no direct link between $s$ and $d$, \textit{i.e.} all the transmitted information should be forwarded by the relays. 
    To perform the proposed security method, $s$ and $d$  should share the CSI knowledge of the source-trusted
relay-destination link, which is the kernel of our developed security method. This CSI is considered to be the main cause of randomness and it 
 is completely mapped into a vector $V$ of digital values.
 It should be noted that this security algorithm is implemented just before the communication process starts, and it can be renewed at any time $s$ and $d$ agree on to keep refreshing the source of security and to make it as strong as possible. 
In the first hop, the source will encode the data prior to the transmission by using the vector $V$. Then, $s$ will use this vector again to map its transmission of different messages $x_i$'s towards  $N$ sectors, where  $\mathop 1\le i \le \mathop N$, $
\mathop N \in  \mathop \mathbb{N}\nolimits^+$. 
We will denote the desired encoded signal by $x_{tr}$, whereas the other signals $\mathop x\nolimits_{i \ne {tr}}$ are the fake ones that are transmitted over the other sectors.
Without the knowledge of $V$, each untrusted relay $e$ will try to randomly guess the useful signal with a probability {\color{black}{$1/N$}}. Even if it succeeds in guessing and receiving the useful message, the untrusted relay would still need the vector $V$ to decode it. In the second phase, all the $K$ relays will resend their received messages towards $d$ using DBF technique.  
Since it has the same vector $V$, the destination will be able to know from which sector the useful message is coming and decode it using $V$. 
The received signal expression, at a $k$th  relay, where $1 \le k \le K$, is given by \vspace{-3pt} \begin{eqnarray}
\mathop y\nolimits_k  =  \sqrt {\mathop P\nolimits_i } \mathop h\nolimits_{s,k}\mathop x\nolimits_i  +   \mathop n\nolimits_k ,
\vspace{-6pt}
\end{eqnarray}
where $n_k\sim\mathcal{N}_c\left( {0,\mathop \sigma \nolimits^2 } \right)$ is the complex additive white Gaussian noise (AWGN)  at the $k$th relay, with mean 0 and variance $\sigma^2$, $P_i$ is  the transmitted power from $s$ towards the $i$th sector.
 We assumed that the channels are quasi-static block log-normal channels, \textit{i.e.} the channel coefficient $h_{v,r} \sim \operatorname{ln\mathcal{N}}(\mu_v, \sigma_v^2)$, where $\{v,r\}\subset \{\{s,k\},\{k,d\}\}$, is considered as constant during the transmission time of one message, but it may change independently  thereafter, the CSI  is known by the receiving nodes, and the noise variance $N_0$ has the same value in the first and the second phases.
 It is important to note  that adopting such security solution by allowing a part of the nodes to forward fake messages is feasible due to the availability of a high number of electronic sensors with limited capabilities.
 Consequently, the received signal-to-noise ratio (SNR), at a $k$th relay, is expressed as
\vspace{-3pt}
\begin{eqnarray}\label{Gammak111}
\mathop \gamma \nolimits_{k}  ={\mathop \rho \nolimits_s \mathop {\left| {\mathop h\nolimits_{s,k} } \right|}\nolimits^2 },
\end{eqnarray}
where  $
\mathop \rho \nolimits_j \mathop  = \limits^\Delta  {{\mathop P\nolimits_j } \mathord{\left/
 {\vphantom {{\mathop P\nolimits_j } {\mathop N\nolimits_0 }}} \right.
 \kern-\nulldelimiterspace} {\mathop N\nolimits_0 }}$, $j \in \left\{ {s,k,e} \right\}$. In the second hop, the retransmitted message from the $k$th relay will be
  $\mathop \chi \nolimits_k = \mathop \alpha \nolimits_k \mathop w\nolimits_k \mathop y\nolimits_k 
$, 
where $w_k$ is the beamforming weight, and $\mathop \alpha \nolimits_k$ represents the normalized amplifying coefficient
$
\mathop \alpha \nolimits_k  = \frac{1}{{\sqrt {\mathop \rho \nolimits_s \mathop {\left| {\mathop h\nolimits_{s,k} } \right|}\nolimits^2  +   \mathop N\nolimits_0 } }}.
$
The received useful messages at $d$ will be written as
\begin{eqnarray}\label{yoverd}
\mathop y\nolimits_d  = \sum\nolimits_{m = 1}^M {\mathop {\mathop h\nolimits_{m,d} \alpha }\nolimits_m \mathop w\nolimits_m \mathop y\nolimits_m }  + \mathop n\nolimits_d ,
\end{eqnarray}
{\color{black}{where $M$ is the number of the relays in the sector that receives the right message.}} $1 \le m\le M$, $n_d\sim\mathcal{N}_c\left( {0,N_0 } \right)$ is the complex AWGN at  $d$.
After \hspace{-1pt}optimizing \hspace{-1pt}the beamforming\hspace{-1pt} weights from \hspace{-1pt}\cite{7061484} and \hspace{-1pt}the\hspace{-1pt} references\hspace{-1pt} therein,\hspace{-1pt} the\hspace{-1pt} SNR \hspace{-1pt}at\hspace{-1pt} the \hspace{-1pt}destination \hspace{-1pt}is \hspace{-1pt}obtained \hspace{-1pt}as
\begin{eqnarray}\label{Gammad}
\mathop \gamma \nolimits_d  = \sum\nolimits_{m = 1}^M {\frac{{\mathop \rho \nolimits_s \mathop {\left| {\mathop h\nolimits_{s,m} } \right|}\nolimits^2 \mathop \rho \nolimits_m \mathop {\left| {\mathop h\nolimits_{m,d} } \right|}\nolimits^2 }}{{\mathop \rho \nolimits_s \mathop {\left| {\mathop h\nolimits_{s,m} } \right|}\nolimits^2    + \mathop \rho \nolimits_m \mathop {\left| {\mathop h\nolimits_{m,d} } \right|}\nolimits^2  + 1}}=\sum\nolimits_{m = 1}^M\gamma_m.} 
\end{eqnarray}
The channel capacity at $d$ will be \vspace{-3pt} \begin{eqnarray}\label{Cd}
\mathop C\nolimits_d  = \mathop {\left[ {\frac{1}{2}\log \left( {1 + \mathop \gamma \nolimits_d } \right)} \right]}\nolimits^ +, 
\vspace{-3pt}
\end{eqnarray}
where $[\xi \mathop ]\nolimits^ +$ denotes $\max \{ \xi ,0\}$. 
\subsection{Non colluding eavesdropping relays}
In this scenario, there are two different hypotheses $H1$ and $H2$ as follows
: \\
Hypothesis $H_1$: the untrusted  relay is in the right sector with a probability {\color{black}{$\mathop p_1= 1/N$}} and it knows how to recover $V$ and decode the message.\\
Hypothesis $H_2$: the untrusted relay is in a wrong sector,   with a probability $\mathop p_0=1-\mathop p_1={\color{black}{ 1-1/N}}$. Then, this relay will not impact the security and the channel capacity at the eavesdropper $e$ will be equivalent to zero from a security point of view.
Considering the aforementioned two hypotheses, the channel capacity at $e$ will be expressed as \vspace{-3pt}
\begin{eqnarray}\label{equiva34}
\mathop C\nolimits_e  = \left\{ {\begin{array}{*{20}{c}}
{\frac{1}{2}\log \left( {1 + \mathop \gamma \nolimits_e } \right)}&{\mathop H\nolimits_1 }\\
0&{\mathop H\nolimits_2,}
\end{array}} \right.
\end{eqnarray}
where $
\mathop \gamma \nolimits_e  = \mathop \rho \nolimits_s \mathop {\left| {\mathop h\nolimits_{s,e} } \right|}\nolimits^2
$ is the SNR of the useful message at $e$. 
\subsection{ colluding eavesdropping relays}
Assuming aggressive untrusted relays, cooperating between each other and sending their messages towards an external wire-tapper $A$, the received useful signal at $A$ will be written as
\vspace{-3pt}
\begin{eqnarray}
\mathop y\nolimits_A  = \sum\nolimits_{u = 1}^{U_1} {\mathop {\mathop h\nolimits_{{u},A} \mathop\alpha\nolimits_{u} } \mathop w\nolimits_{u} \mathop y\nolimits_{u} }  + \mathop n\nolimits_A ,
\end{eqnarray}
where $U_1$ is the number of the untrusted relays that are in the right sector and  sending the useful messages $x_{tr}$, and $1 \le {u}\le {U_1} \le U$. Moreover, $n_A\sim\mathcal{N}_c\left( {0,N_0 } \right)$ is the complex AWGN at  $A$.
Hence, the SNR at $A$ will become\vspace{-3pt}
\begin{eqnarray}\label{gamagame}
\mathop \gamma \nolimits_A  \hspace{-1pt}=\hspace{-4pt}\sum\nolimits_{u = 1}^{\mathop {U}\nolimits_1 } {{\frac{{\mathop \rho \nolimits_s \mathop {\left| {\mathop h\nolimits_{s,u} } \right|}\nolimits^2 \mathop \rho \nolimits_{u} \mathop {\left| {\mathop h\nolimits_{u,A} } \right|}\nolimits^2 }}{{\mathop \rho \nolimits_s \hspace{-1pt}\mathop {\left| {\mathop h\nolimits_{s,u} } \right|}\nolimits^2  \hspace{-1pt}  +\hspace{-1pt} \mathop \rho \nolimits_{u}\hspace{-1pt} \mathop {\left| {\mathop h\nolimits_{u,A} } \right|}\nolimits^2 \hspace{-1pt} +\hspace{-1pt} 1}}} }\hspace{-1pt}=\hspace{-4pt}\sum\nolimits_{{u} = 1}^{\mathop U\nolimits_1 } {\mathop \gamma \nolimits_{\mathop u} }.
\end{eqnarray}
We will define two hypotheses for $A$: \\
Hypothesis $\mathop H\nolimits_1^{\ensuremath{'}}$: $A$  receives the right message  with a probability {\color{black}{$\mathop p_1=1/N$}}  and knows how to recover $V$ and decodes the message.\\
Hypothesis ${\mathop H\nolimits_2^{\ensuremath{'}}}$: the colluding relays are just in the wrong sectors, or  $A$ con not recover $V$, which means that $A$ won't have any impact on the security.
Hence, the channel capacity at $A$ will be equivalent to\vspace{-2pt}
\vspace{-5pt}
\begin{eqnarray}\label{equiva74}
\mathop C\nolimits_A  = \left\{ {\begin{array}{*{20}{c}}
{\frac{1}{2}\log \left( {1 + \mathop \gamma \nolimits_A } \right)}&{\mathop H\nolimits_1^{\ensuremath{'}} }\\
0&{\mathop H\nolimits_2^{\ensuremath{'}}},
\end{array}} \right.
\end{eqnarray}
We will define the  worst security case as when $e$, (in the non colluding  state), or $A$, (in the colluding  state),  knows how to recover $V$ and decode the message. Therefore, the channel capacity at $q$, where $q \in \{ e, A\}$, is given as
\vspace{-3pt}
\begin{eqnarray}\label{zizoyoyo}
C_q    =
\mathop {\left[ {\color{black}{\frac{1}{N}}}.\frac{1}{2}\log \left( {1 + {\mathop \gamma \nolimits_{\mathop q } } } \right) \right]}\nolimits^ +.
\end{eqnarray}
 From \eqref{Cd}  and \eqref{zizoyoyo}, the general secrecy capacity expression of the worst case is calculated as
 \vspace{-3pt}
\begin{eqnarray}\label{Cd-Ce}
 \mathop C\nolimits_{S,q}\hspace{-1pt}=\hspace{-2pt} \mathop {\left[ {C_d  -C_q  } \right]}\nolimits^ +\hspace{-3pt}=\hspace{-2pt} \mathop {\left[ {\frac{1}{2}\hspace{-1pt}\log \left( {1\hspace{-2pt} +\hspace{-2pt} \mathop \gamma \nolimits_d } \right) - {\color{black}{\frac{1}{2N}}}\hspace{-1pt}\log \left( {1\hspace{-2pt} +\hspace{-2pt} \mathop \gamma \nolimits_q } \right)} \right]}\nolimits^ +\hspace{-4pt}.
\end{eqnarray}  
\section{Secrecy Outage Probability}\label{soppur}
\begin{theorem}\label{theorem1j}
The secrecy outage probability expression of our proposed method $C_{S,q}$, for both passive and aggressive untrusted relays scenarios, is expressed as
\[
\Pr \left[ {\mathop C\nolimits_{S,q}  < R} \right] 
=\frac{2}{3}\Phi\left(\left( {{{\ln \left( {\mathop 2\nolimits^{2R} \mathop {\left( {1 + \mathop e\nolimits^{\mathop \mu \nolimits_q } } \right)}\nolimits^{{\color{black}{\frac{1}{N}}}}  - 1} \right) - \mathop \mu \nolimits_d }}} \right)\mathop \sigma \nolimits_d^{-1} \right)\]
\[ + \frac{1}{6}\Phi\left(\left( {{{\ln \left( {\mathop 2\nolimits^{2R} \mathop {\left( {1 + \mathop e\nolimits^{\left( {\mathop \mu \nolimits_q  + \sqrt 3 \mathop \sigma \nolimits_q } \right)} } \right)}\nolimits^{{\color{black}{\frac{1}{N}}}}  - 1} \right) - \mathop \mu \nolimits_d }}} \right)\mathop \sigma \nolimits_d^{-1}\right)  \]
\vspace{-13pt}
\begin{eqnarray}\label{SOPeq}
- \frac{1}{6}\Phi\left( \hspace{-2pt}\left(\hspace{-1pt}{{{\ln \hspace{-1pt}\left(\hspace{-1pt} {\mathop 2\nolimits^{2R} \mathop {\left( {1 + \mathop e\nolimits^{\left( {\mathop \mu \nolimits_q  - \sqrt 3 \mathop \sigma \nolimits_q } \right)} } \right)}\nolimits^{{\color{black}{\frac{1}{N}}}}  - 1}\hspace{-2pt} \right)\hspace{-2pt} -\hspace{-1pt} \mathop \mu \nolimits_d }}}\hspace{-2pt}\right)\hspace{-2pt} \mathop \sigma \nolimits_d^{-1}\hspace{-2pt}\right)\hspace{-2pt}.
\end{eqnarray}
\end{theorem}
\begin{IEEEproof}
From \eqref{Cd-Ce}, and for a threshold $R$, the SOP is defined as \cite{7061484}\vspace{-9pt}
\begin{eqnarray}\label{PrCsr}
\Pr \left[ {\mathop C\nolimits_{S,q}  \hspace{-1pt}<\hspace{-2pt} R} \right] \hspace{-1pt}=\hspace{-1pt} \Pr \hspace{-1pt}\left[ {\frac{1}{2}\hspace{-1pt}\log \hspace{-1pt}\left( {1\hspace{-2pt} +\hspace{-2pt} \mathop \gamma \nolimits_d } \right) - {\color{black}{\frac{1}{2N}}}\hspace{-1pt}\log \hspace{-1pt}\left( {1\hspace{-2pt} +\hspace{-2pt} \mathop \gamma \nolimits_q } \right)\hspace{-1pt} <\hspace{-2pt} R} \right]
\end{eqnarray}
\[
 = \Pr \left[ {\mathop \gamma \nolimits_d  < \mathop 2\nolimits^{2R} \mathop {\left( {1 + \mathop \gamma \nolimits_q } \right)}\nolimits^{{\color{black}{\frac{1}{N}}}}  - 1} \right]
\]
\[\vspace{-3pt}
 = \int\nolimits_0^\infty  {\mathop F\nolimits_{\mathop \gamma \nolimits_d } \left( {\mathop 2\nolimits^{2R} \mathop {\left( {1 + \mathop \gamma \nolimits_q } \right)}\nolimits^{{\color{black}{\frac{1}{N}}}}  - 1} \right)\mathop f\nolimits_{\mathop \gamma \nolimits_q } \left( {\mathop \gamma \nolimits_q } \right)d\mathop \gamma \nolimits_q }.
\]
 Since $\gamma_q$ and $\gamma_d$ are  following a log-normal distribution, (please refer to Appendix A for the proof), then their probability density function (PDF) and  cumulative distribution function (CDF)
 are given as follows
\vspace{-9pt}
\begin{equation}\label{pdf1}
f_X(x;\mu,\sigma) = \frac{1}{ x\sigma \sqrt{2 \pi}}\, e^{-\frac{(\ln x - \mu)^2}{2\sigma^2}},
\end{equation}
\vspace{-9pt}
\begin{eqnarray}\label{cdf1}
F_X(x;\mu,\sigma)  
= \Phi\bigg(\frac{\ln x - \mu}{\sigma}\bigg),
\end{eqnarray}
 and $\Phi$ is the CDF of the standard normal distribution. 
Thus, the SOP in \eqref{PrCsr} is obtained as
\vspace{-3pt}
\begin{eqnarray}\label{PrCsr2}
\Pr \left[ {\mathop C\nolimits_{S,q}  < R} \right] =
\end{eqnarray}
\[
 = \int\limits_0^\infty  {\Phi\left( {\frac{{\ln \left( {\mathop 2\nolimits^{2R} \mathop {\left( {1 + \mathop \gamma \nolimits_q } \right)}\nolimits^{{\color{black}{\frac{1}{N}}}}  - 1} \right) - \mathop \mu \nolimits_d }}{{\mathop \sigma \nolimits_d }}} \right)\frac{{\mathop e\nolimits^{ - \frac{{\mathop {\left( {\ln \mathop \gamma \nolimits_q  - \mathop \mu \nolimits_q } \right)}\nolimits^2 }}{{2\mathop {\mathop \sigma \nolimits_q }\nolimits^2 }}} }}{{\mathop \gamma \nolimits_q \mathop \sigma \nolimits_q \sqrt {2\pi } }}d\mathop \gamma \nolimits_q }.
\]
\vspace{-3pt}
\begin{eqnarray}\label{betad}
\text{Let} \hspace{+10pt}\beta = \ln \left( {\mathop \gamma \nolimits_q } \right) ,\hspace{6pt}\text{then} \hspace{6pt}\mathop \gamma \nolimits_q  = \mathop e\nolimits^\beta  ,\hspace{6pt}\text{and}\hspace{6pt}    d\mathop \gamma \nolimits_q  = \mathop e\nolimits^\beta  d\beta.
\end{eqnarray} 
 $\beta$ is a normally distributed r.v. $\beta \sim \operatorname{\mathcal{N}}(\mu_q, \sigma^2_q).$ 
Substituting \eqref{betad} in \eqref{PrCsr2}, the secrecy outage probability  
$\Pr \left[ {\mathop C\nolimits_{S,q}  < R} \right]$ is written as
\vspace{-9pt}
\begin{eqnarray}\label{expectationres1}
\int\limits_0^\infty  {\overbrace {\Phi\left( {\frac{{\ln \left( {\mathop 2\nolimits^{2R} \mathop {\left( {1 + \mathop e\nolimits^\beta } \right)}\nolimits^{{\color{black}{\frac{1}{N}}}}  - 1} \right) - \mathop \mu \nolimits_d }}{{\mathop \sigma \nolimits_d }}} \right) }^{\psi \left( \beta \right)}\frac{{\mathop e\nolimits^{ - \frac{{\mathop {\left( {\beta - \mathop \mu \nolimits_q } \right)}\nolimits^2 }}{{2\mathop {\mathop \sigma \nolimits_q }\nolimits^2 }}} }}{{\mathop \sigma \nolimits_q \sqrt {2\pi } }}d\beta }. 
\end{eqnarray}
It is noticed that \eqref{expectationres1} denotes  the expectation of ${\psi \left(\beta \right)}$. We will use \textit{Holtzman}  tool \cite{135712} to approximate $ E\left[ {\psi \left( \beta \right)} \right]$  in terms of three points located at $\mu_q$, $\mathop \mu \nolimits_q  + \sqrt 3 \mathop \sigma \nolimits_q $ and $\mathop \mu \nolimits_q  - \sqrt 3 \mathop \sigma \nolimits_q $ as follows\vspace{-1pt}
\[
\Pr \left[ {\mathop C\nolimits_{S,q}  < R} \right] = E\left[ {\psi \left( \beta \right)} \right] = \]
\vspace{-17pt}
\begin{eqnarray}\label{zaza31}
\frac{2}{3}\psi \left( {\mathop \mu \nolimits_q } \right) + \frac{1}{6}\psi \left( {\mathop \mu \nolimits_q  + \sqrt 3 \mathop \sigma \nolimits_q } \right) - \frac{1}{6}\psi \left( {\mathop \mu \nolimits_q  - \sqrt 3 \mathop \sigma \nolimits_q } \right).
\end{eqnarray}
Compensating ${\psi \left( \beta \right)}$ from \eqref{expectationres1} in \eqref{zaza31} yields \eqref{SOPeq}.
\end{IEEEproof}
{{
\section{Simulation Results}
\begin{figure}
\centering
\includegraphics[trim =48 150 00 230, clip,width=3.6in]{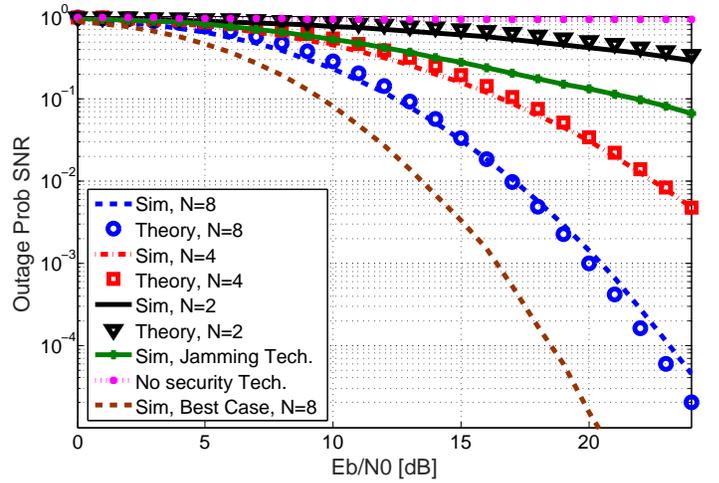}
\caption{SOP with passive untrusted relays: $R=3$ bps/Hz, $M=4$, $\sigma_s=\sigma_k=0.95$ and $\mu_s=\mu_k=1$.
}
\label{wiretap model2}
\end{figure}
\begin{figure}
\centering
\includegraphics[trim =10 160 00 191, 
clip,width=3.5in]{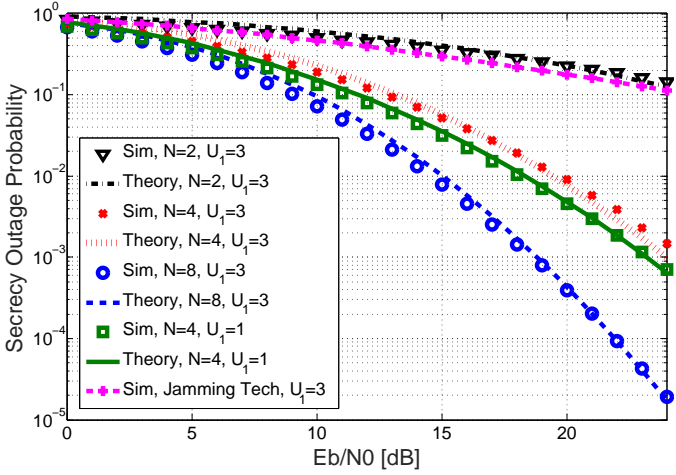}
\caption{SOP with aggressive untrusted relays: $R=2$ bps/Hz, $M=4$, $\sigma_s=\sigma_k=1.1$ and $\mu_s=\mu_k=0.69$.
}
\label{wiretap model3}
\end{figure}
In this section, we demonstrate the validity of our derived results {\color{black}{ using MATLAB software}}.
Fig.\ref{wiretap model2} shows the SOP as a function of the SNR. It is noticed that the derived expressions accurately characterize the simulation results. It is assumed that $R=3$ bps/Hz, $M=4$, $\sigma_s=\sigma_k=0.95$ and $\mu_s=\mu_k=1$. 
From Fig.\ref{wiretap model2}, we can see how the secrecy performance improves when the number of sectors $N$ is increased. For example, to keep the SOP level  at $10^{-2}$, the source has to increase the number of sectors $N$ from $4$ to $8$, which will also reduces the required SNR from  $23 dB$ to $17 dB$.
Also, it is shown that our performed technique outperforms the conventional jamming technique, {\color{black}{where the destination jams the nodes while the source is transmitting in the first hop}}.
As we can see from Fig.\ref{wiretap model2}, the margin between the worst and the best case, when $e$ does not know how to recover $V$,  depends on $e$'s capability in recovering $V$ and decoding the message.

Fig.\ref{wiretap model3} shows the SOP of our proposed technique for different values of $N$, when  $R=2$ bps/Hz, $\sigma_s=\sigma_k=1.1$, and $\mu_s=\mu_k=0.69$. 
It can be seen that the greater the number of sectors, the better the secrecy performance. 
Moreover, we can see from Fig.\ref{wiretap model3} that there is not that much of difference between the base of one and that of three aggressive untrusted relays. For example, at SNR level of $18 dB$, the SOP just goes from $1.05\times 10^{-2}$ to $1.85\times 10^{-2}$ after adding two extra aggressive untrusted relays, which means that our proposed technique is immune towards adding more eavesdropping relays that are  cooperating with each other. 
Also, it is shown that the security performance is improved when our technique is applied compared to the jamming technique. 
\section{Conclusions}
In this paper, we proposed a new location-based multicasting protocol that is mapped by the knowledge of a trusted link's  CSI in two-hops WSN. 
We provided an analytical study for the SOP for the passive and the aggressive behaviors of the untrusted relays. The results showed the immunity of our technique towards the untrusted relays aggressive behavior, and an improvement in the security compared to the conventional jamming technique. 
\appendices
\section{ }
We will prove that $\gamma_q$ and $\gamma_d$ are  following a log-normal distribution. First, we will define the  SNR $\mathop \gamma \nolimits_{i,j}$ as follows \vspace{-7pt}
\begin{eqnarray}\label{Gammaij}
\mathop \gamma \nolimits_{i,j}  = \mathop \rho \nolimits_i \mathop {\left| {\mathop h\nolimits_{i,j} } \right|}\nolimits^2 , \vspace{-9pt}
\end{eqnarray} \vspace{-4pt}
 where $i\in \left\{ {s,m} \right\}$ and $j \in \left\{ {m,e,d} \right\}$.  \begin{lem1}\label{Lem3}
 Let $X \sim \operatorname{\ln\mathcal{N}}(\mu, \sigma^2)$, then $aX\hspace{-3pt} \sim\hspace{-1pt} \operatorname{\ln\mathcal{N}}( \mu \hspace{-1pt}+ \ln a,\ \sigma^2),\,\,$ and $X^a \sim \operatorname{\ln\mathcal{N}}( a\mu, a^2\ \sigma^2),\,\,           a\in \mathbb{R}.$
\end{lem1} \vspace{-4pt}
\vspace{-3pt}
From \textbf{Lemma \ref{Lem3}}, where $a =2$, the channel gain 
 $\mathop {\left| {\mathop h\nolimits_{\mathop i,j} } \right|}\nolimits^2 \sim \operatorname{\ln\mathcal{N}}(2\mu_{\gamma_{i,j}}, 4\sigma^2_{\gamma_{i,j}})$. 
By using the properties in  \textbf{Lemma \ref{Lem3}}, we find that  
 $\gamma_{i,j} \sim \operatorname{\ln\mathcal{N}}(\mu_{\gamma_{i,j}}, \sigma^2_{\gamma_{i,j}})$, where 
$\mu_{\gamma_{i,j}}=2\mathop \mu \nolimits_i  + \ln \left( {\mathop \rho \nolimits_i } \right)$, $\ln \left( {\mathop \rho \nolimits_i } \right)=\ln \left( {\mathop P \nolimits_i } \right)-\ln \left( {\mathop N\nolimits_0 } \right)$, and $ \sigma^2_{\gamma_{i,j}}=4\mathop {\mathop \sigma \nolimits_i }\nolimits^2$. Hence, $\gamma_{e} \sim \operatorname{\ln\mathcal{N}}(\mu_{\gamma_{s,e}}, \sigma^2_{\gamma_{s,e}})$.

Now, we will find the distribution of $\gamma_{m}$   \eqref{Gammad} with the following approximation for high SNRs, as follows 
\vspace{-3pt}
\begin{eqnarray}\label{Gammak}
\mathop \gamma \nolimits_{m}\hspace{-2pt}  = \hspace{-2pt}\frac{{\mathop \gamma \nolimits_{s,m} \mathop \gamma \nolimits_{m,d} }}{{\mathop \gamma \nolimits_{s,m} \hspace{-2pt} +\hspace{-2pt} \mathop \gamma \nolimits_{m,d} \hspace{-1pt} +\hspace{-1pt} 1}}\mathop \approx  \frac{{\mathop \gamma \nolimits_{s,m} \mathop \gamma \nolimits_{m,d} }}{{\mathop \gamma \nolimits_{s,m} \hspace{-1pt} + \hspace{-1pt}\mathop \gamma \nolimits_{m,d} }}\hspace{-2pt} = \hspace{-2pt}\frac{1}{{ {\frac{1}{{\mathop \gamma \nolimits_{s,m} }}} \hspace{-2pt}+\hspace{-2pt}  {\frac{1}{{\mathop \gamma \nolimits_{m,d} }}}}} \hspace{-2pt}= \hspace{-2pt}\frac{1}{z},
\end{eqnarray}
\vspace{-4pt}
where  $z =  \mathop z\nolimits_1  + \mathop z\nolimits_2$, 
$z_1=\frac{1}{{ \gamma_{s,m} }}$  and $z_2=\frac{1}{{ \gamma_{m,d} }}$.
\begin{lem1}\label{Lem4}
 Let $X_j \sim \operatorname{\ln\mathcal{N}}(\mu_j,\sigma_j^2)\ $ are independent log-normally distributed variables with varying $\sigma$ and $\mu$ parameters, and $Y=\textstyle\sum_{j=1}^n X_j$. Then the distribution of $Y$ has no closed form expression, but can be reasonably approximated by another log-normal distribution $Z$ with parameters\cite{MyListOfPapers:Fenton1960}
 \vspace{-1pt}
 \begin{eqnarray}\label{muz}
  \mu_Z &= \ln\!\left[ \sum e^{\mu_j+\sigma_j^2/2} \right] - \frac{\sigma^2_Z}{2},
  \end{eqnarray}
 \vspace{-16pt}
 \begin{eqnarray}\label{sigmaz}
  \sigma^2_Z &= \ln\!\left[ \frac{\sum e^{2\mu_j+\sigma_j^2}(e^{\sigma_j^2}-1)}{(\sum e^{\mu_j+\sigma_j^2/2})^2} + 1\right].
  \end{eqnarray}
\end{lem1}
Form \textbf{Lemma \ref{Lem3}}, where $a=-1$, we find that  $Z_1 \sim \operatorname{\ln\mathcal{N}}(-\mu_{\gamma_{s,m}}, \sigma^2_{\gamma_{s,m}})$ and 
$Z_2 \sim \operatorname{\ln\mathcal{N}}(-\mu_{\gamma_{m,d}}, \sigma^2_{\gamma_{m,d}})$. Also, from \textbf{Lemma \ref{Lem4}},  
  $Z \sim \operatorname{\ln\mathcal{N}}(\mu_z, \sigma^2_z)$, where \vspace{-3pt} 
\[
\mathop {\mathop \sigma \nolimits_z }\nolimits^2  = \ln \left( {\left( {\exp \left( {\mathop \sigma \nolimits_{\mathop z\nolimits_1 }^2 } \right) - 1} \right)/2 + 1} \right),
\]
\[
\mathop \mu \nolimits_z  = \ln \left( {2\exp \left( {\mathop \mu \nolimits_{\mathop z\nolimits_1 } } \right)} \right) + 0.5\left( {\mathop \sigma \nolimits_{\mathop z\nolimits_1 }^2  - \mathop \sigma \nolimits_z^2 } \right).
\vspace{-3pt}
\]
Thus, from \textbf{Lemma \ref{Lem3}} and \eqref{Gammak}, we get $\gamma_{m} \sim \operatorname{\ln\mathcal{N}}(\mu_{\gamma_{m}}, \sigma^2_{\gamma_{m}})$, where $a=-1$,  
$\mathop \mu \nolimits_{\mathop \gamma \nolimits_{m} }  =  - \mathop \mu \nolimits_z$, and  $\mathop \sigma \nolimits_{\mathop \gamma \nolimits_{m} }^2  = \mathop \sigma \nolimits_z^2 
$. 
 From \eqref{Gammad}, since $\gamma_d$ is a sum of many $\gamma_{m}$, we will again use  \textbf{Lemma \ref{Lem4}}
 to find that 
$\gamma_d \sim \operatorname{\ln\mathcal{N}}(\mu_{\gamma_d}, \sigma^2_{\gamma_d})$, where\vspace{-3pt}
\[\vspace{-3pt}
\mathop {\mathop \sigma \nolimits_d }\nolimits^2  = \ln \left( {\left( {\exp \left( {\mathop \sigma \nolimits_{\mathop \gamma \nolimits_{m} }^2 } \right) - 1} \right)/M + 1} \right),
\]
\[\vspace{-3pt}
\mathop \mu \nolimits_d  = \ln \left( {M\exp \left( {\mathop \mu \nolimits_{\mathop \gamma \nolimits_{m} } } \right)} \right) + 0.5\left( {\mathop \sigma \nolimits_{\mathop \gamma \nolimits_{m} }^2  - \mathop \sigma \nolimits_d^2 } \right).
\]
Since the expressions of $\gamma_{u}$   and  $\gamma_A$ in  \eqref{gamagame} are similar to $\gamma_m$ and $\gamma_d$ respectively, by following the same steps, we show that  $\gamma_{u} \sim \operatorname{\ln\mathcal{N}}(\mu_{\gamma_{m}}, \sigma^2_{\gamma_{m}})$ and  
$\gamma_{A} \sim \operatorname{\ln\mathcal{N}}(\mu_{\gamma_{A}}, \sigma^2_{\gamma_{A}})$ where
\vspace{-3pt}
\[
\vspace{-6pt}
\mathop {\mathop \sigma \nolimits_{A} }\nolimits^2  = \ln \left( {\left( {\exp \left( {\mathop \sigma \nolimits_{\mathop \gamma \nolimits_{m} }^2 } \right) - 1} \right)/{U_1} + 1} \right),
\]
\vspace{-12pt}
\[
\vspace{-17pt}
\mathop \mu \nolimits_{A}  = \ln \left( {{U_1}\exp \left( {\mathop \mu \nolimits_{\mathop \gamma \nolimits_{m} } } \right)} \right) + 0.5\left( {\mathop \sigma \nolimits_{\mathop \gamma \nolimits_{m} }^2  - \mathop \sigma \nolimits_{A}^2 } \right).
\]  
\vspace{-3pt}
}}
\bibliographystyle{IEEEtran}
\bibliography{IEEEabrv,MyListOfPapers}
\end{document}